\documentclass[manuscript]{aastex}
\usepackage{xcolor}

\fullcollaborationName{The Friends of AASTeX Collaboration}

\begin{document}

\title{On the onset delays of solar energetic electrons and protons: Evidence for a common accelerator}

%% Use \author, \affil, plus the \and command to format author and affiliation 
%% information.  If done correctly the peer review system will be able to
%% automatically put the author and affiliation information from the manuscript
%% and save the corresponding author the trouble of entering it by hand.
%%
%% The \affil should be used to document primary affiliations and the
%% \altaffil should be used for secondary affiliations, titles, or email.

%% Authors with the same affiliation can be grouped in a single
%% \author and \affil call.
\author{R.D. Strauss\altaffilmark{1,2}, N. Dresing\altaffilmark{3}, I.G. Richardson\altaffilmark{4,5}, J.P. van den Berg\altaffilmark{1}, P.J. Steyn\altaffilmark{1}}

%% Notice that each of these authors has alternate affiliations, which
%% are identified by the \altaffilmark after each name.  Specify alternate
%% affiliation information with \altaffiltext, with one command per each
%% affiliation.

\altaffiltext{1}{Center for Space Research, North-West University, Potchefstroom, 2522, South Africa}
\altaffiltext{2}{National Institute for Theoretical and Computational Sciences (NITheCS), South Africa} 
\altaffiltext{3}{Department of Physics and Astronomy, University of Turku, Turku, Finland}
\altaffiltext{4}{NASA/Goddard Space Flight Center, Greenbelt, MD 20771, USA}
\altaffiltext{5}{Department of Astronomy, University of Maryland, College Park, MD 20742, USA}

%% Mark off the abstract in the ``abstract'' environment. 
\begin{abstract}

The processes responsible for the acceleration of solar energetic particles (SEPs) are still not well understood, including whether SEP electrons and protons are accelerated by common or separate processes. Using a numerical particle transport model that includes both pitch-angle and perpendicular spatial diffusion, we simulate, amongst other quantities, the onset delay for MeV electrons and protons and compare the results to observations of SEPs from widely-separated spacecraft. Such observations have previously been interpreted, in a simple scenario assuming no perpendicular diffusion, as evidence for different electron and proton sources.  We show that, by assuming a common particle source together with perpendicular diffusion, we are able to simultaneously reproduce the onset delays for both electrons and protons. We argue that this points towards a common accelerator for these particles. Moreover, a relatively broad particle source is required in the model to correctly describe the observations. This is suggestive of diffusive shock acceleration occurring at large shock structures playing a significant role in the acceleration of these SEPs.

\end{abstract}


%%%%%%%%%%%%%%%%%%%%%%%%%%%%%%%%%%%%%%%%%%%%%%%%%%%%%%%%%%%%%%%%%%%%%%
\section{Introduction}

The potential acceleration mechanisms for solar energetic particles (SEPs) are still heavily debated. The most likely acceleration regions are magnetic reconnection (or a closely related mechanism) in solar flares and shock acceleration (either shock drift acceleration or diffusive shock acceleration) related to shocks driven by coronal mass ejections (CMEs). SEP observations at Earth have led to the (potentially oversimplified) classification of events into impulsive (sometimes also referred to as electron-rich), that are associated with short-duration solar flares, and gradual SEP events, that are associated with CMEs and long-duration flares \citep[e.g.][]{Reames2013}, though some studies have suggested that there is a continuum of SEP event properties \citep[e.g.][]{canetal2010}. It is therefore not clear whether SEP electrons and protons are accelerated by the same acceleration mechanism during the same transient solar events. \\

Previous work has shown a clear empirical relationship between MeV electron and proton measurements. For instance, the \citet{Posner2007} Relativistic Electron Alert System for Exploration (REleASE) algorithm uses an empirically established relationship between relativistic electron intensities and subsequent proton intensities to predict future proton levels. More recently \citet{Dresingetal2022} found a similar dependence of electron and proton intensities on shock parameters suggesting a common shock-related accelerator for both species. {Additionally, \citet{Richardsonetal2014} found clear linear (in logarithmic space) relationships between onset and peak delays at the observing spacecraft relative to the onset of the related type III radio emission for different levels of magnetic connection between the spacecraft and the associated flare location. However, these linear relationships from \citet{Richardsonetal2014} are different for electrons and protons and cannot be explained by simple ballistic motion between a single expanding source and the observer, as also discussed by \citet{Kollhoffetal2021} in relation to the widespread SEP event on November 29, 2020. If particle transport is neglected, the different electron and proton dependencies can be explained by different sources for these SEP electrons and protons, expanding in longitude at different rates.} \\

In this work we examine whether SEP electron and proton measurements at Earth can in fact be explained by a common acceleration source when interplanetary transport (i.e. particle scattering) is also considered. The focus is on simulating the particle onset delays as presented by \citet{Richardsonetal2014}. The modelling approach applied here includes perpendicular (cross-field) diffusion. There is increasing evidence that perpendicular diffusion is an essential transport process for SEPs. \citet{Kouloumvakosetal2022}, for instance, show that spacecraft that are magnetically unconnected to a CME-driven shock can still observe a significant SEP increase. The low particle anisotropy levels associated with such poorly connected observers \citep[e.g.][]{Dresingetal2014} are also consistent with significant levels of cross-field diffusion: {As the perpendicular diffusion process is generally much slower than parallel transport, the initial, highly anisotropic, beam of SEPs predominantly propagates along the mean field without being scattered significantly away from it. However, perpendicular diffusion becomes increasingly effective at later times during the nearly-isotropic decay phase of the SEP events. These nearly-isotropic particle distributions, being scattered effectively perpendicular to the mean field later in the SEP event, are usually associated with low levels of particle anisotropy.} Additionally, modelling by e.g. \citet{Drogeetal2016} has shown that significant levels of perpendicular diffusion is needed to reproduce many observed SEP events. \\

%%%%%%%%%%%%%%%%%%%%%%%%%%%%%%%%%%%%%%%%%%%%%%%%%%%%%%%%%%%%%%%%%%%%%%
\section{The numerical transport model}

\begin{figure*}[t]
    \centering
\includegraphics[width=0.99\textwidth]{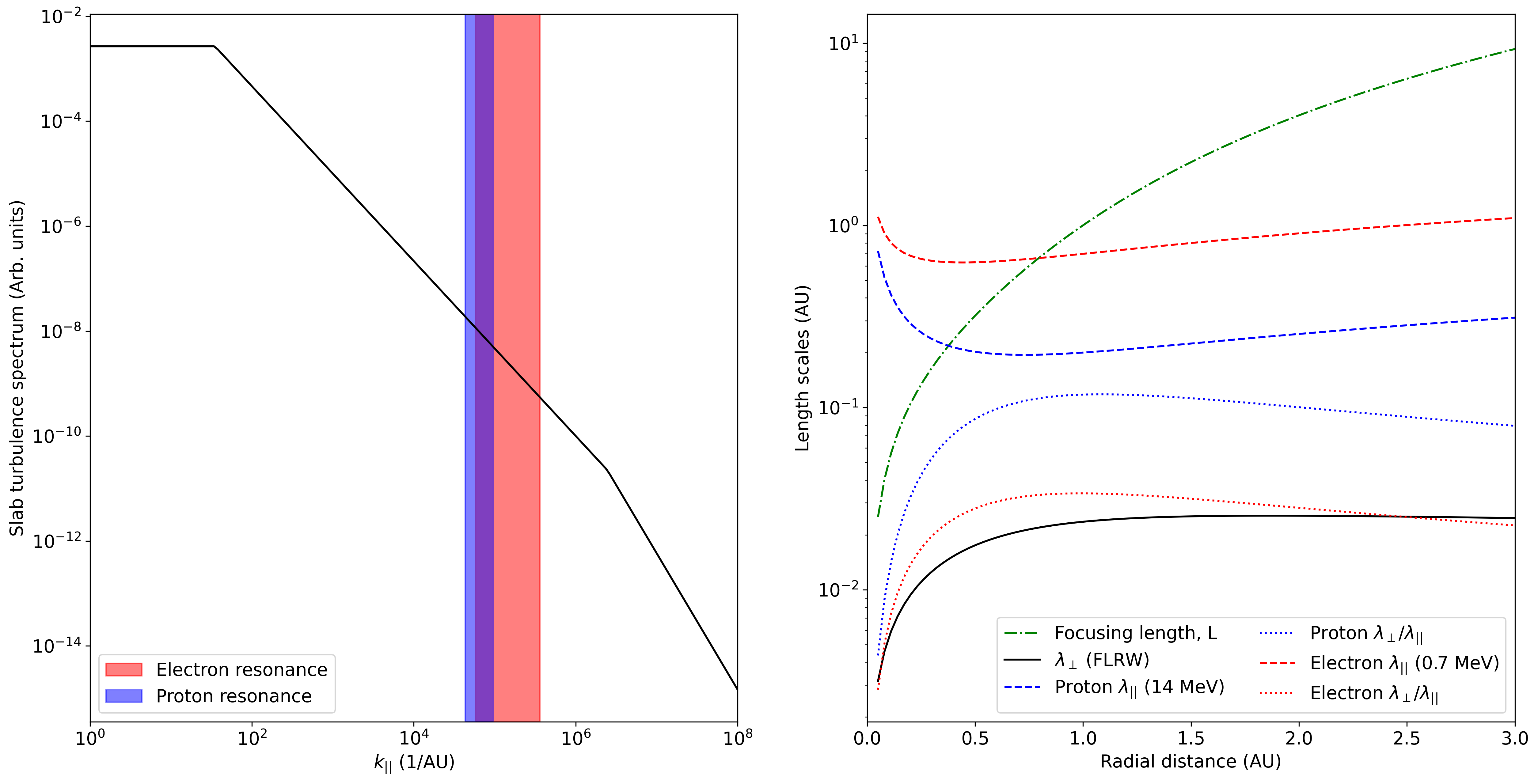}
    \caption{{The left panel shows the assumed slab turbulence spectrum at Earth, as a function of the parallel wavenumber, while the shaded regions indicate where electrons (red shading) and protons (blue shading) in the assumed energy ranges will resonate}. {The right panel shows the} resulting electron and proton parallel and perpendicular mean-free-paths as used in this study. Also shown is the magnetic focusing length.}
    \label{fig:lambdas}
\end{figure*}

In this work we simulate the transport of SEPs through the turbulent interplanetary medium by solving the following focused transport equation for the distribution function, $f$,

\begin{eqnarray}
\label{Eq:TPE}
\frac{\partial f}{\partial t} + \nabla \cdot \left( \mu v \hat{b} f \right) + \frac{\partial}{\partial \mu} \left( \frac{1-\mu^2}{2L} vf \right) &=&   
 \frac{\partial}{\partial \mu} \left(D_{\mu\mu}  \frac{\partial f}{\partial \mu} \right)\nonumber \\  
&+&   \nabla \cdot \left( \mathbf{D}^{(x)}_{\perp}\cdot \nabla f \right) 
\end{eqnarray}

using the approach outlined by \citet{StraussFichtner2015}. Here, $\mu$ is the particle pitch-angle cosine with respect to the mean magnetic field, directed in the $\hat{b}$ direction \citep[see e.g.][]{vandenbergetal2020}. This model was previously applied to near-relativistic electron transport by \citet{Straussetal2017} and \citet{Straussetal2020}. In these previous applications, the focus was on calculating the pitch-angle diffusion coefficient, $D_{\mu\mu}$, and perpendicular diffusion coefficient, $D_{\perp}$, from first principles using observed solar wind turbulence values. We continue with that approach here, but apply the model to both SEP electron and proton transport. {Although we have tried, as far as possible, to constrain all the turbulence quantities from measurements, we accept that there are still some uncertainties related to these quantities and the associated diffusion coefficients calculated from them}. {The left panel of Fig. \ref{fig:lambdas} shows the assumed slab turbulence spectrum, as a function of the parallel wavenumber $k_{||}$, at Earth. The red and blue shaded regions indicate where electrons and protons in the assumed energy ranges will, respectively, resonante. Here we use the energy ranges of \citet{Richardsonetal2014}, namely 0.7 -- 4 MeV for electrons and 14 -- 24 MeV for protons. For this illustrative example we assume that the particles will resonante at $k_{||} \sim r_L^{-1}$, where $r_L$ is the particles' Larmor radius \citep[][]{Straussetal2020}. This calculation illustrates that protons and electrons, in the assumed energy ranges, will resonate in the inertial range of the slab turbulence, with resonate wavenumbers very close to each other}. As usual, a \cite{Parker1958} heliospheric magnetic field (HMF) geometry is assumed with an associated focusing length, $L$. This value, along with the resulting parallel and perpendicular mean-free-paths for electrons and protons, using the calculations of \citet{Straussetal2017}, are shown, as a function of radial distance, in {the right panel of} Fig. \ref{fig:lambdas}. The perpendicular diffusion coefficient is discussed in more detail in the next section. {We do not include drift effects as these are generally considered to be negligible for low energy (i.e. $\sim$ MeV) particles \citep[][]{Engelbrechetetal2017,vandenbergetal2021}.}\\

As an inner boundary condition ($r_0 = 0.05$ AU) to the model, the following function is specified

\begin{eqnarray}
    f(r=r_0, \phi,t) = \frac{C}{t} \exp \left[ - \frac{\tau_a}{t}  - \frac{t}{\tau_e} \right] \exp \left[ - \frac{\left( \phi - \phi_0 \right)}{2 \sigma^2}  \right].
\end{eqnarray}

The time dependence of this function is determined by the so-called acceleration ($\tau_a = 1$ hr) and escape ($\tau_e = 1$ hr) timescales, while a Gaussian source (in terms of longitude) is specified where the broadness, $\sigma$, can be varied. This function quantifies the assumed accelerated SEP distribution released into the interplanetary medium from the particle accelerator. With the uncertainty surrounding the acceleration process and the fact that these timescales cannot be directly measured, these values are subject to change in future simulations. While $\tau_e$ seemingly has only a small effect on the simulated intensities, the calculated peak delay (discussed later) is sensitive to the choice of $\tau_a$. Work is under way to further constrain these model inputs.  \\

\subsection{Efficiency of perpendicular diffusion}

As with previous work we assume that perpendicular diffusion is governed by the so-called field line random walk (FLRW) process \citep[][]{Jokipii1966}, leading to a perpendicular diffusion coefficient of the form

\begin{equation}
    \label{Eq:FLRW}
    D_{\perp} = a |\mu| v  \kappa_{FL} ,
\end{equation}

where $\kappa_{FL}$ is the so-called FLRW diffusion coefficient which describes the diffusion of the magnetic field lines and depends on the underlying turbulence quantities \citep[see also][]{Straussetal2016}. Note that $D_{\perp}$ is proportional to particle speed and that, in general, faster particles diffuse faster perpendicular to the mean HMF. The factor $a$ is introduced to account for the fact that particles have a finite gyro-radius and cannot perfectly follow diffusing field lines \citep[e.g.][]{Shalchi2009}. The efficiency of perpendicular diffusion in this model therefore strongly depends on the seemingly {\it ad hoc} factor $a$ which can thus be interpreted as {\it the probability that particles are tied to fluctuating magnetic field lines} \citep[e.g.][]{QinShalchi2014}. While some progress has been made to derive $a$ from first principles \citep[][]{Shalchi2015,Shalchi2020}, in previous work, \citet{Straussetal2017} treated $a$ as a free parameter and obtained a good comparison between SEP measurements, for 100 keV electrons, and modelling results when $a=0.2$. Although this value is based on a comparison with a single particle species at a single energy, we adopt $a=0.2$ for all simulations performed in this work. \\

%%%%%%%%%%%%%%%%%%%%%%%%%%%%%%%%%%%%%%%%%%%%%%%%%%%%%%%%%%%%%%%%%%%%%%
\section{Analytical Estimates and Model Results}

%%%%%%%%%%%%%%%%%%%%%%%%%%%%%%%%%%%%%%%%%%%%%%%%%%%%%%%%%%%%%%%%%%%%%%
\subsection{Analytical Estimates}

\begin{figure}[t]
    \centering
\includegraphics[width=0.99\textwidth]{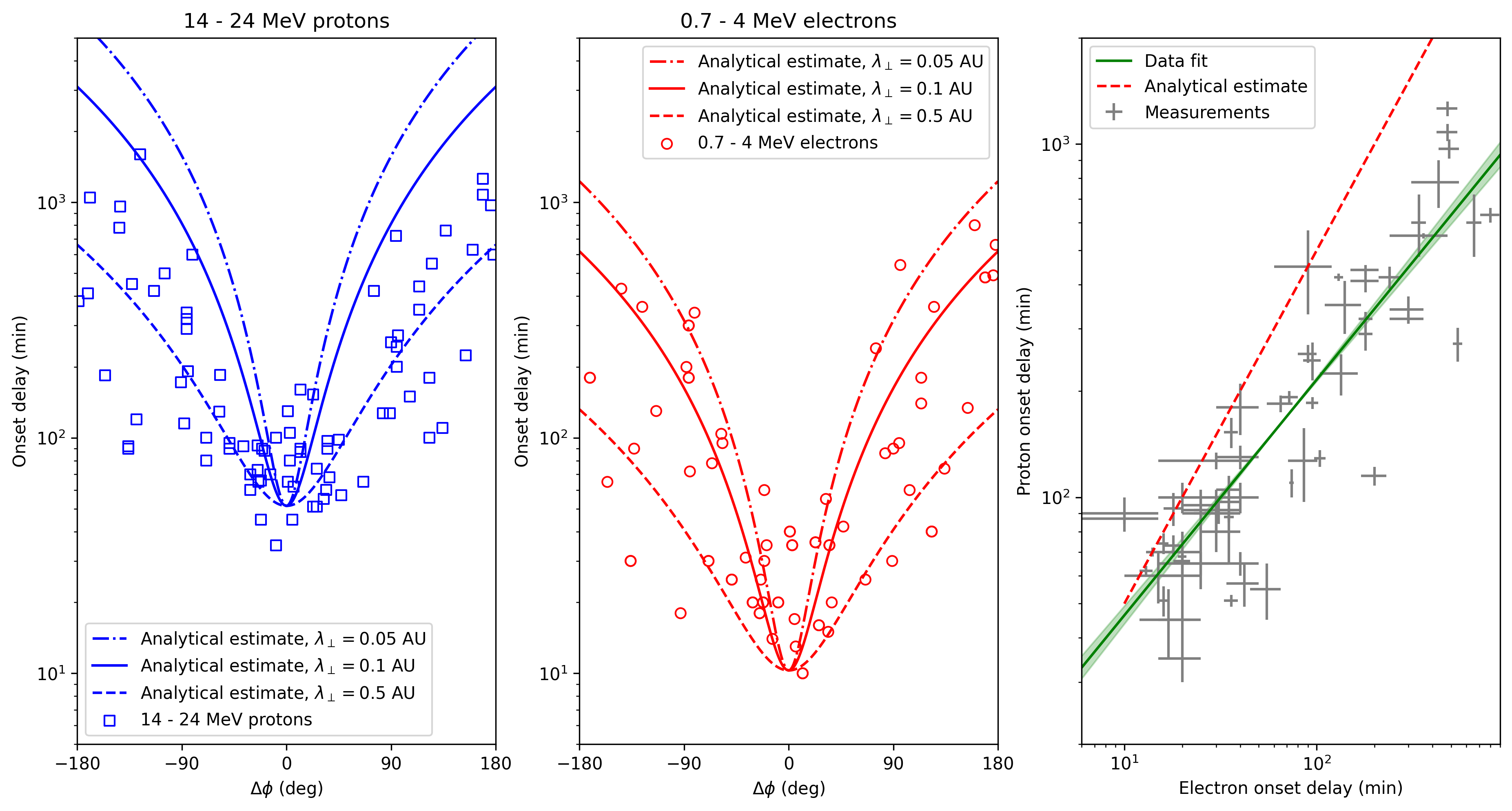}
    \caption{The left and middle panels are analytical estimates of the onset delay, for a SEP point source, as a function of magnetic connectivity for protons and electrons, respectively. The right panel shows the relationship between the electron and proton onset delays which is independent of $\lambda_{\perp}$. Observations are taken from \citet{Richardsonetal2014}, while analytical estimates are shown for different assumptions of $\lambda_{\perp}$.}
    \label{fig:analytical_estimates}
\end{figure}

In this section we start by estimating the onset delay as a function of magnetic connection using simplistic analytical arguments. For an observer perfectly magnetically connected to the SEP source, the SEPs simply have to propagate along the mean HMF. Assuming this happens in a ballistic (scatter free) fashion we can estimate the onset delay as $\tau_0 = \Delta s_{||}/v$ with $\Delta s_{||} = 1.2$ AU the distance the SEPs would cover along a nominal spiral magnetic field line to  1 AU and $v$ the particle speed. For 1.7 MeV electrons this is $v_e \sim 7$ AU/hr and for 18 MeV protons, $v_p \sim 1.4$ AU/hr. If we now assume that particles propagate diffusively across HMF lines to reach magnetically unconnected observers, and that this process takes an additional time $\tau_d$, the onset delay becomes $\tau = \tau_0 + \tau_d$. We can estimate $\tau_d$ by assuming diffusive motion perpendicular to the mean HMF is due to perpendicular diffusion. For an isotropic distribution (implying significant scattering), the diffusive propagation time can be evaluated as $\tau_d \sim \left(\Delta s_{\perp} \right)^2/ \left( 6 \kappa_{\perp} \right)$ \citep[see e.g.][]{StraussPotgieter2011}. The isotropic form of the perpendicular diffusion coefficient is

\begin{equation}
    \kappa_{\perp} = \frac{1}{2} \int_{-1}^{+1} D_{\perp} (\mu) d\mu = \frac{a v}{2}  \kappa_{FL},
\end{equation}

and, in terms of the perpendicular mean-free-path, $\lambda_{\perp} = 3 \kappa_{\perp}/v = (3/2) a \kappa_{FL}$. Note that $\lambda_{\perp}$ for the FLRW has no speed (energy) dependence and is determined completely by the underlying magnetic turbulence.  If we assume the distance covered perpendicular to the field, $\Delta s_{\perp}$, is only directed in the azimuthal direction (i.e. neglecting the non-radial geometry of the HMF), we can estimate $\Delta s_{\perp} \sim \Delta s_{||} \Delta \phi$ where $\Delta \phi$ is the azimuthal angle away from the best magnetic connection to the source. The resulting onset delay then becomes

\begin{equation}
    \tau = \frac{\Delta s_{||}}{v} \left[ 1 + \frac{\Delta s_{||} }{2 \lambda_{\perp}} \left( \Delta \phi \right)^2 \right].
\end{equation}

These estimates are shown for electrons and protons in Fig. \ref{fig:analytical_estimates} for different values of $\lambda_{\perp}$. The left and middle panels show the calculated onset delay as a function of magnetic connectivity for protons (left) and electrons (middle). The right panel shows the resulting relationship between the electron and proton onset delays, $\tau_p = (v_e/v_p) \tau_e$ that only depends on the ratio of the particles' speeds in this approach. \\

In this simplified derivation we assume ballistic motion parallel to the field, but diffusive motion of an isotropic distribution perpendicular to the mean HMF. The effect of pitch-angle scattering and potentially anisotropic distributions are therefore not included consistently and must be included in a modelling approach to obtain a correct estimate of the onset delay. This will be done in the following section where a more realistic Parkerian geometry, {that is not simply a radially directed magnetic field,} is also included. However, even with this very simplified approach we note that a combination of motion along the mean HMF and diffusion perpendicular to the mean field is sufficient to explain the observed dependence of the onset delay on magnetic connection. Additionally, by assuming a FLRW-type perpendicular diffusion coefficient (which is itself proportional to particle speed) we can explain, to first order, the linear relationship between the observed electron and proton onset delays. A more detailed calculation is performed in the next section.\\

%%%%%%%%%%%%%%%%%%%%%%%%%%%%%%%%%%%%%%%%%%%%%%%%%%%%%%%%%%%%%%%%%%%%%%
\subsection{On the Geometry of Perpendicular Diffusion}
\label{Sec:geomatry}

\begin{figure*}[t]
    \centering
\includegraphics[width=0.49\textwidth]{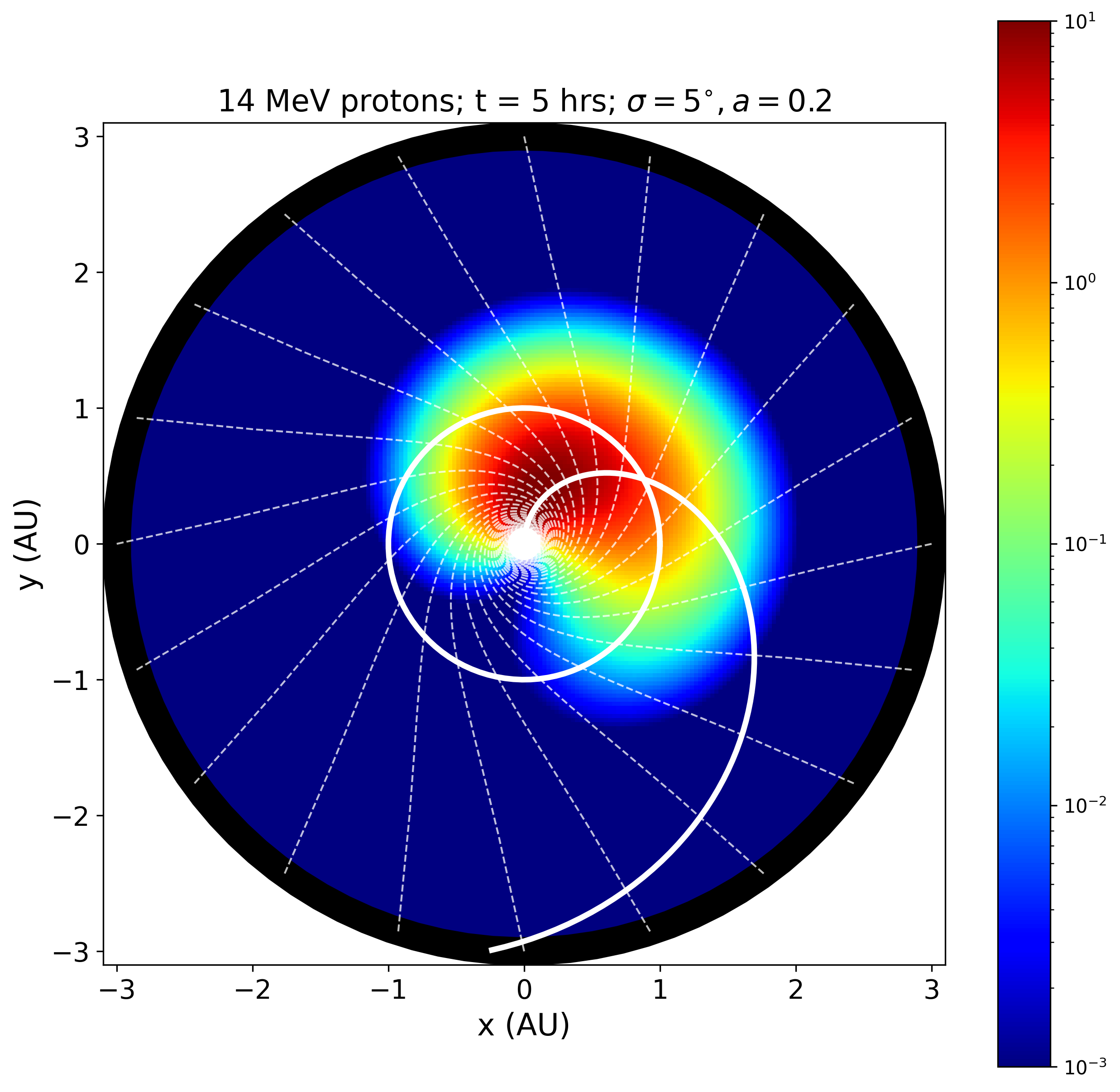}
\includegraphics[width=0.49\textwidth]{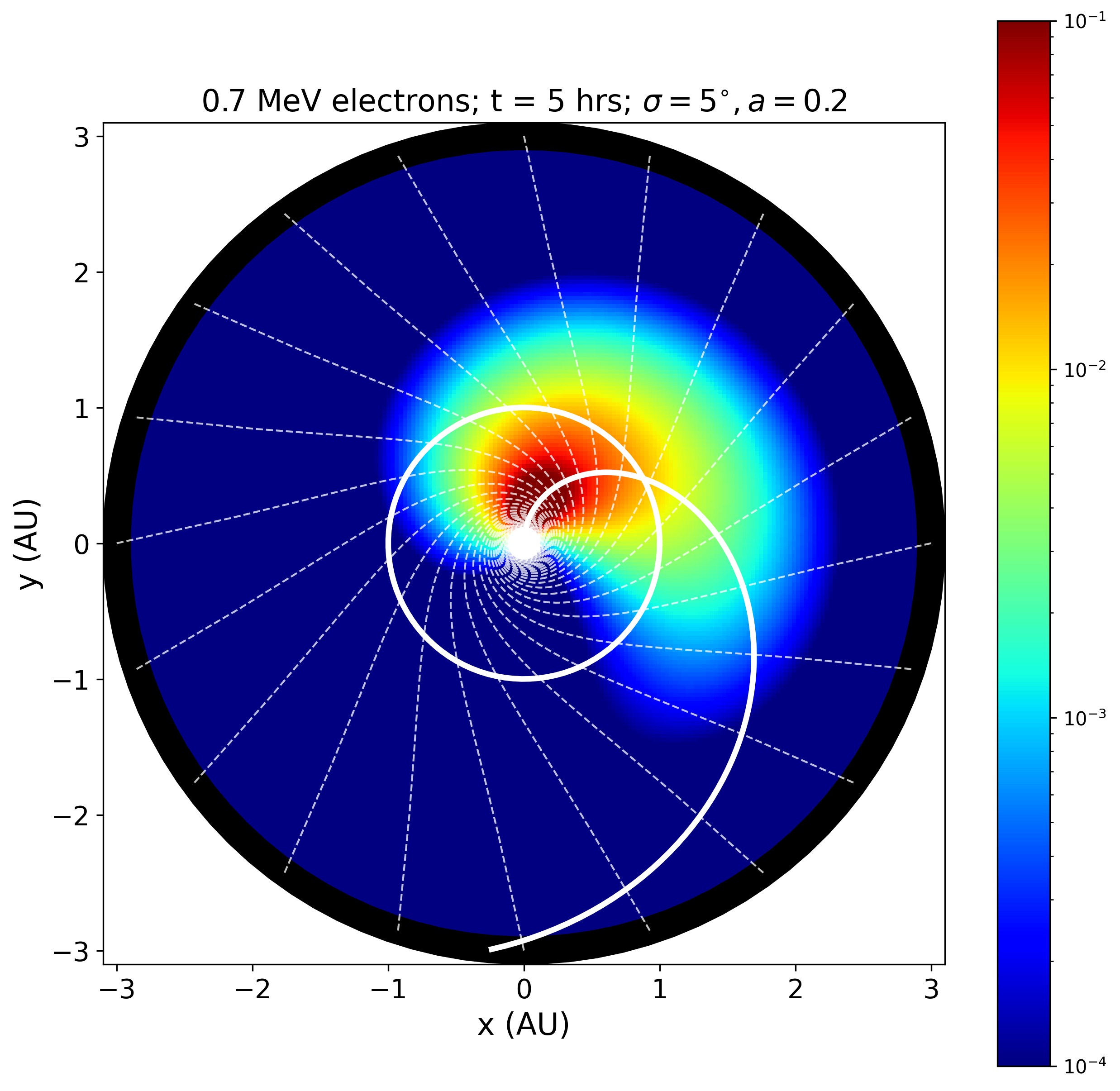}
    \caption{Contour plots of the SEP proton (left) and electron (right) intensity 5 hrs after particle injection.}
    \label{fig:contour_plots}
\end{figure*}

Assuming a narrow source with $\sigma = 5^{\circ}$, Fig. \ref{fig:contour_plots} shows contour plots of the normalized intensity of 14 MeV protons (left panel) and 0.7 MeV electrons (right panel) 5 hrs after the initial particle release. The orbit of Earth (circular at a radius of 1 AU) is shown, along with the HMF line connected to the center of the SEP injection, as thick white lines. Also included, as thin dotted white lines, are line elements perpendicular to the HMF at different radial positions. Perpendicular diffusion acts to transport particles along these line elements. It is important to keep in mind that perpendicular diffusion does not imply motion purely in the longitudinal direction, but owing to the Parker HMF geometry (the HMF spiral angle is $\sim 45^{\circ}$ at Earth's orbit), particles are also transported in the radial direction. For a fixed source this usually implies more efficient particle transport towards the west of the best-connected HMF line, {as these particles are transported {\it away} from the Sun}, leading to asymmetrical distributions in terms of longitude. {In this context, see also the recent simulations of \citet{Laitinenetal2023}.} \\

%%%%%%%%%%%%%%%%%%%%%%%%%%%%%%%%%%%%%%%%%%%%%%%%%%%%%%%%%%%%%%%%%%%%%%
\subsection{Longitudinal Dependence of SEP Intensity}

\begin{figure*}[t]
    \centering
\includegraphics[width=0.99\textwidth]{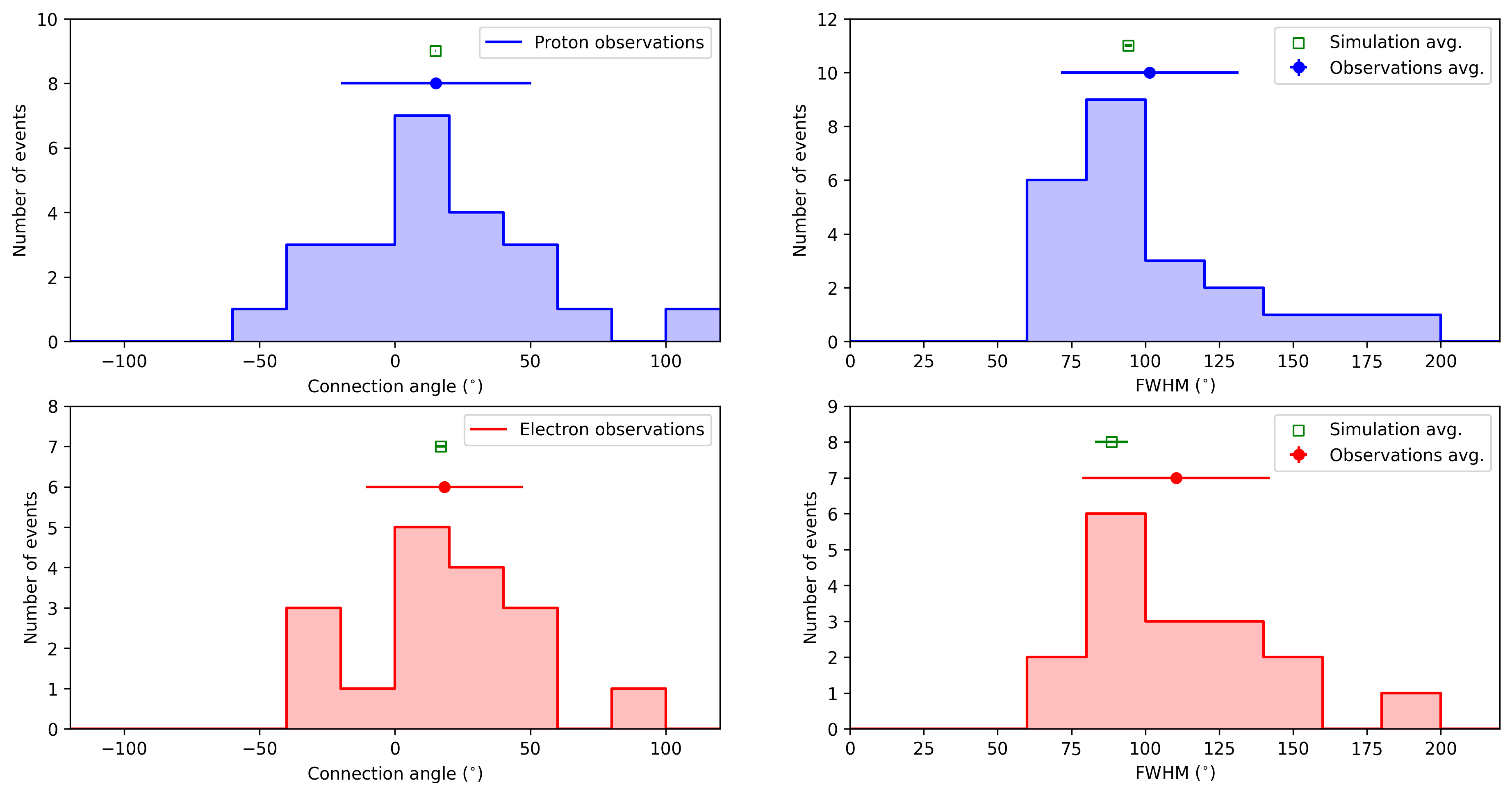}
    \caption{The histograms show the observed distributions of the connection angle (left panels) and full-width half maximum (FWHM; right panels) of 14 -- 24 MeV protons (top panels; in blue) and 0.7 -- 4 MeV electrons (bottom panels; in red) from  Fig. 24 of \citet{Richardsonetal2014}. Observational averages are indicated by the filled circles, while the open green symbols show the results from the best-fit numerical model. }
    \label{fig:histograms}
\end{figure*}

\begin{figure*}[t]
    \centering
\includegraphics[width=0.49\textwidth]{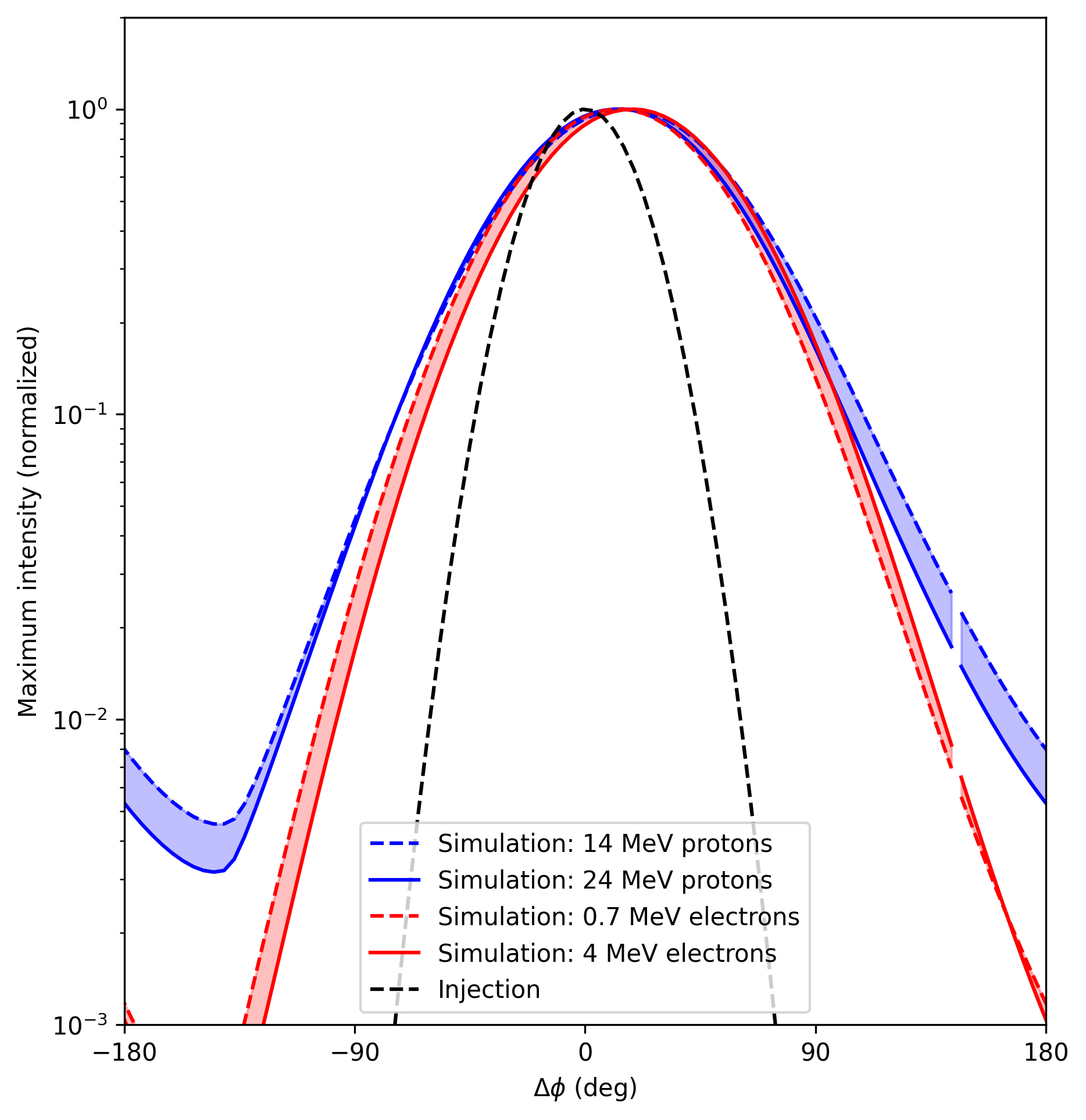}
\includegraphics[width=0.49\textwidth]{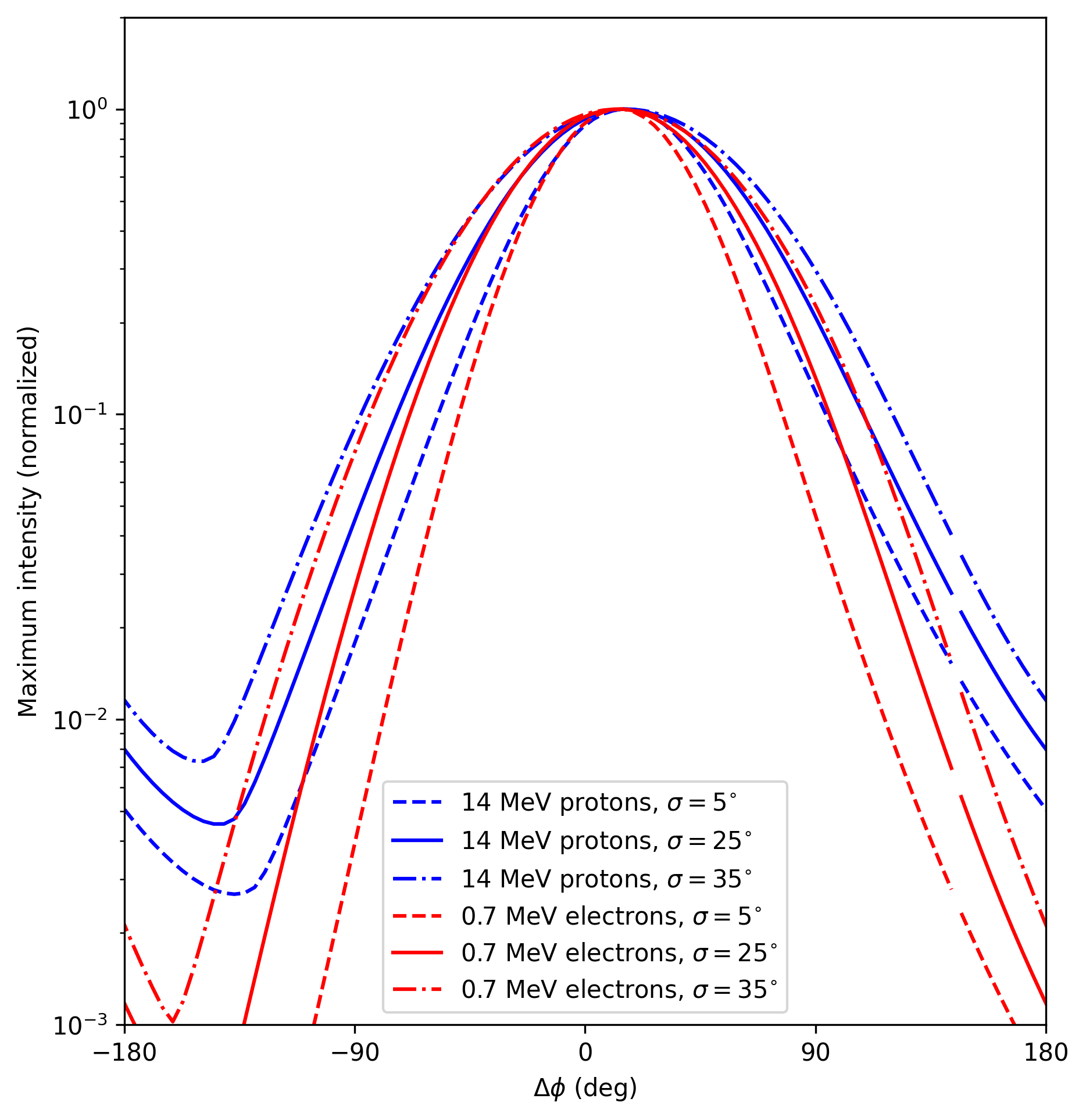}
    \caption{The left panel shows the maximum omni-directional intensity as a function of the magnetic connection angle for best-fit model assuming a source width of $\sigma = 25^{\circ}$ indicated by the black dashed curve. The right panel shows the resulting intensities for different assumptions of the source size (indicated in the legend).}
    \label{fig:peak_intensities}
\end{figure*}

\citet{Richardsonetal2014} examined the longitudinal dependence of the intensity of multi-spacecraft SEP proton and electron events in the energy ranges of 14 -- 24 MeV and 0.7 -- 4 MeV, respectively. This was done by fitting Gaussian distributions to the peak intensities at the different observing spacecraft (STEREO A/B and SOHO). The histograms in Fig. \ref{fig:histograms} show these results: The top (blue) distributions are for protons and the bottom (red) distributions are for electrons. The left panels show the connection angle (i.e. the angular offset between the peak of the Gaussian function and the nominal Parker HMF line connected to the parent active region) with positive values indicating a shift towards the west (i.e. direction of solar rotation). Westward shifts are also inferred in other studies of multi-spacecraft events e.g., \citet{Larioetal2014}, \citet{Cohenetal2017}, and \citet{BrunoRiochardson2021}. The right panels show the width of the Gaussian function in terms of the full-width half maximum (FWHM). All the distributions show inter-event variation which is most likely due to different interplanetary conditions during each event and the difference in the accelerator of each event, e.g. larger or smaller CMEs. While we accept that each SEP event will be different, we aim in this work to rather reproduce the {\it average} characteristics of these events. These averages are indicated above the histograms in Fig. \ref{fig:histograms} as solid circles with an associated error bar. \\

The only {remaining} free parameter in our modelling approach is the size of the particle source, $\sigma$. We now simulate SEP intensities for electrons with energies of 0.7 and 4 MeV, and protons with energies of 14 and 25 MeV, the limits of the energy ranges considered by \citet{Richardsonetal2014}, calculate the maximum omni-directional intensity as a function of longitude, and, from this, calculate both the resulting connection angle and FWHM. The size of the assumed source is then adjusted, in increments of $5^{\circ}$, until a good comparison is obtained with the average values of \citet{Richardsonetal2014}. A best fit is obtained by assuming $\sigma = 25^{\circ}$. These modelling results are included in Fig. \ref{fig:histograms} as the green symbols with the associated error bar indicating the deviation for the different energies (i.e. the upper and lower boundaries of the energy channels). Interestingly, both the model and observations indicate a systematic offset of the distribution towards the west (a positive connection angle of $\sim 15^{\circ}$). In the modelling approach this can be explained by the geometry of the perpendicular diffusion process, as discussed in Sec. \ref{Sec:geomatry}, leading to more effective (perpendicular) transport in this direction. For a best fit value of $\sigma = 25^{\circ}$, the left panel of Fig. \ref{fig:peak_intensities} shows the maximum omni-directional intensity as a function of connection angle for the different particle species of different energies, while the black dashed line shows the assumed source size. It is clear that perpendicular diffusion broadens this initial source, while shifting the peak of the distribution towards the west. The right panel shows the resulting distribution for different source sizes. \\

In this section we adjusted the remaining free parameter in our model, the size of the SEP source, until the measured distribution of the omni-directional intensity could be reproduced. However, if just the particle intensity is considered, this could lead to a degeneracy in the modelling approach: Different combinations of particle sources and levels of perpendicular diffusion can reproduce the same distribution at Earth's orbit. Additional observables are needed to uniquely determine all transport parameters. This was done in \citet{Straussetal2017} and we used their value of $a=0.2$ to fix the level of perpendicular diffusion. However, as we will show in the next section, the model set-up with $a=0.2$ and $\sigma = 25^{\circ}$, does also lead to a decent comparison with the onset delays of MeV electrons and protons. We are therefore confident that the model gives a fair representation of the observations.\\

%%%%%%%%%%%%%%%%%%%%%%%%%%%%%%%%%%%%%%%%%%%%%%%%%%%%%%%%%%%%%%%%%%%%%%
\subsection{Electron and Proton Onset Delays}

\begin{figure*}
    \centering
\includegraphics[width=0.99\textwidth]{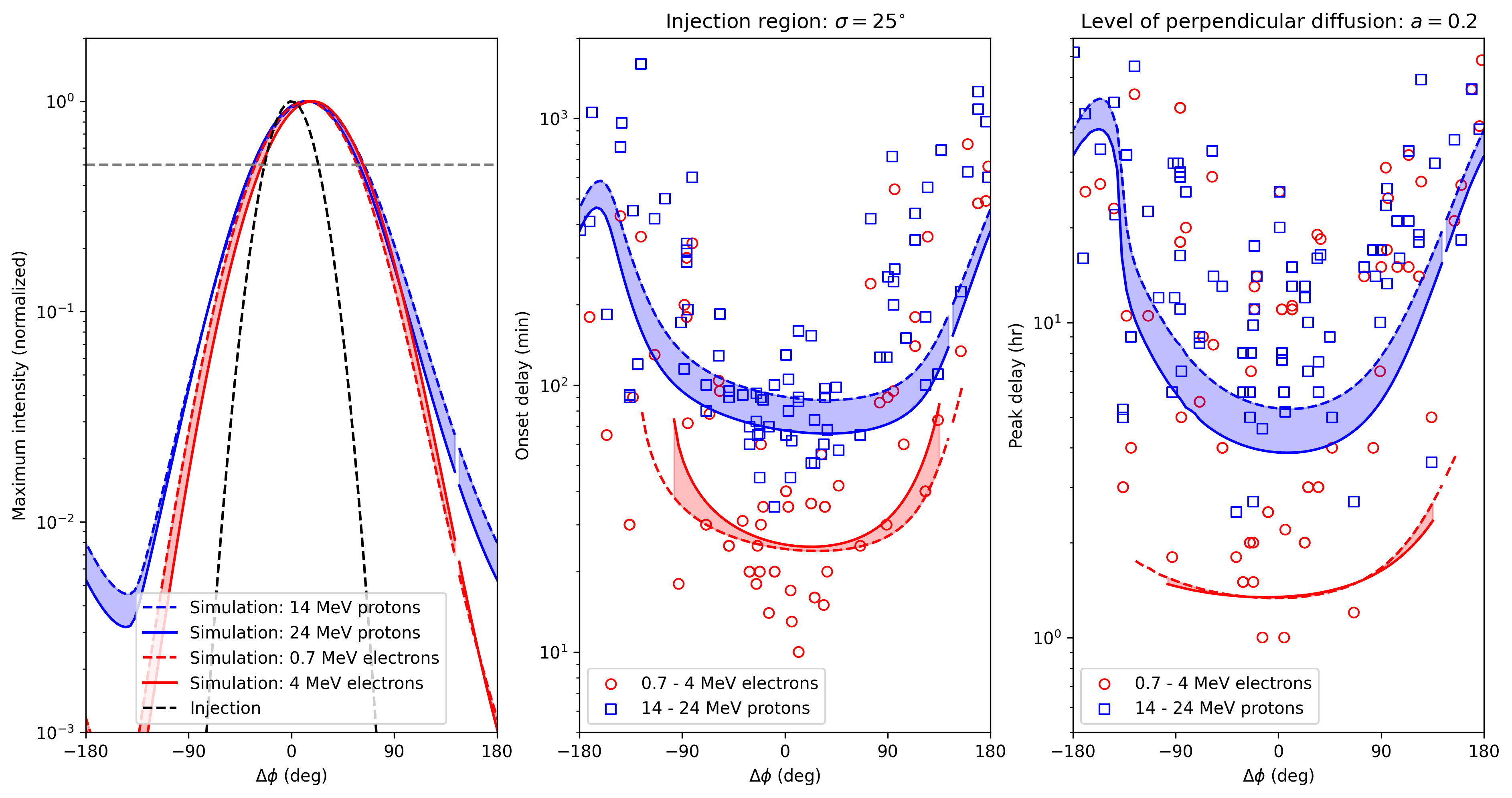}
    \caption{The left panel shows the calculated maximum omni-directional intensity as a function of longitude (i.e. magnetic connection), the middle panel the onset delay in minutes, and the right panel the peak delay in hours. The horizontal dashed line in the left panel indicates 50\% of the peak value. Simulations are performed for electrons and protons of different energies as indicated on the legend. All measurements are taken from \citet{Richardsonetal2014}. }
    \label{fig:onset_times}
\end{figure*}

\begin{figure*}
    \centering
\includegraphics[width=0.99\textwidth]{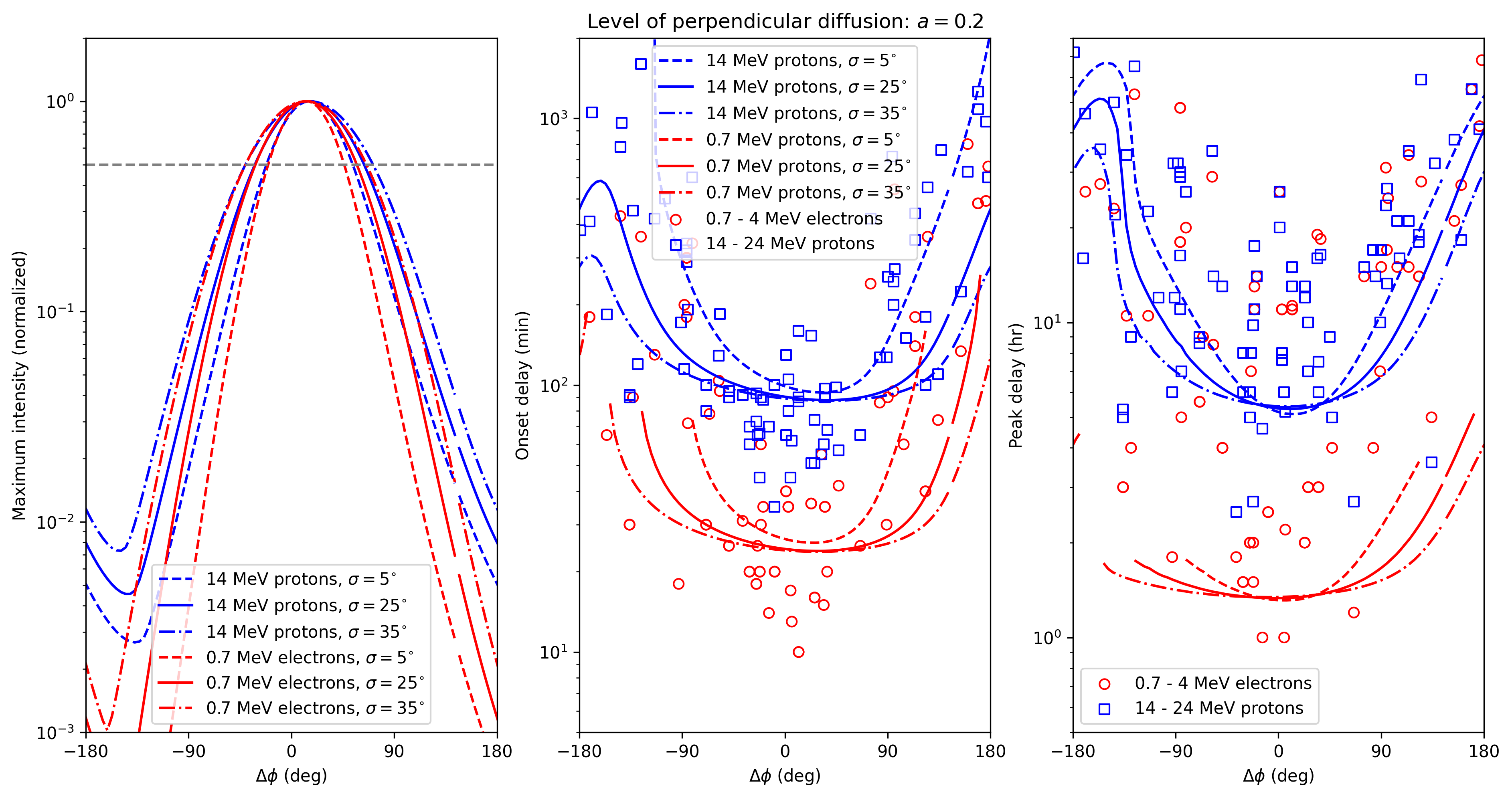}
    \caption{The same as shown in Fig. \ref{fig:onset_times}, but model results are now shown for different source sizes.}
    \label{fig:onset_times_vary_source}
\end{figure*}

\begin{figure*}
    \centering
\includegraphics[width=0.99\textwidth]{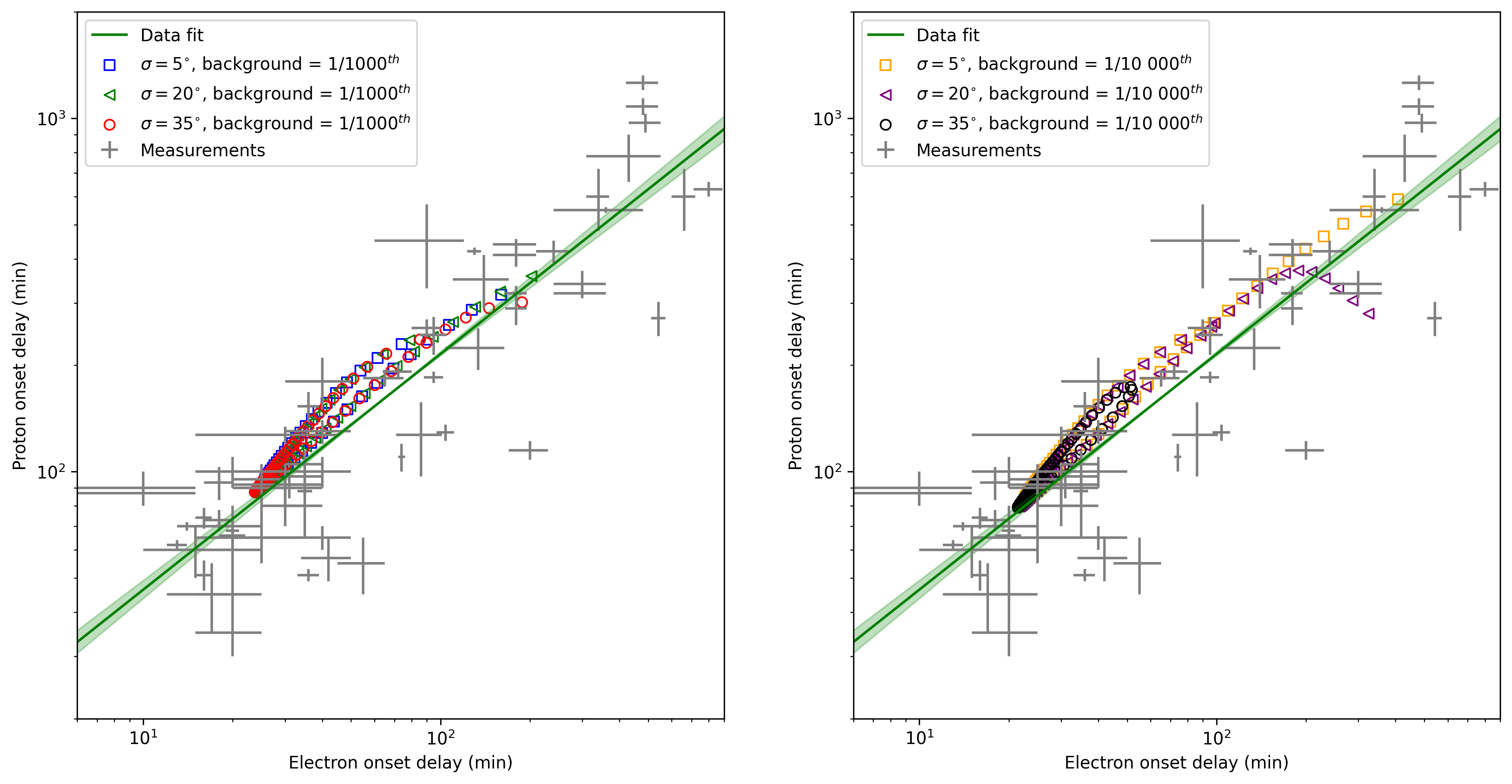}
    \caption{The calculated proton and electron onset delays (the different symbols) compared to the measurements of \citet{Richardsonetal2014}. The left panel shows the model results for different source sizes, while the right panel also assumes a lower background level in the model. }
    \label{fig:onset_times_e_p}
\end{figure*}

In order to model particle onset delays, a relative background needs to be chosen in the model. We follow the previous approach of \citet{Straussetal2017} of calculating the maximum omni-directional intensity, for all longitudes, at 1 AU, and then defining the background as 1/1000$^{th}$ of the peak intensity. The onset delay is then simply calculated as the time from the start of the simulation until this background level is crossed at each longitude. The effects of different background levels are illustrated in Sec. \ref{Sec:caveats}.\\

Using the best-fit model of the previous section, Fig. \ref{fig:onset_times} shows the resulting onset delays. The left panel, once again, shows the maximum omni-directional intensity as a function of longitude (different lines correspond to electrons and protons of different energies as indicated in the legend). The middle panel shows the measurements from \citet{Richardsonetal2014} as the data points (blue again corresponding to protons and red to electrons), while the lines are model calculations. If the modelled intensity, at a given longitude, did not cross the background levels, the onset delay is not defined and no model values are shown. As expected, the minimum onset delay is seen at the best magnetic connection between the observer and the source, where the first arriving (usually very anisotropic) particles reach the observer in an almost scatter-free fashion. The onset delay then increases for increasing magnetic connection due to the relative slow (perpendicular) diffusion process that transports the particles across magnetic field lines. Although the measurements have significant event-to-event variations, the general trend and ball-park values are consistent with the model results.\\

The right panel of Fig. \ref{fig:onset_times} shows the calculated and observed peak delay as a function of longitude. This is calculated as the time that elapsed between particle injection (i.e. the start of the simulation) and the occurrence of the maximum omni-directional intensity at each longitude. Although there is, once again, a fair comparison between model results and observations, the model is unable to reproduce the very long peak delays observed for electrons. This is addressed in more detail in Sec. \ref{Sec:delays}. Fig. \ref{fig:onset_times_vary_source} is similar to Fig. \ref{fig:onset_times} but shows the effect of a changing source size on the model calculations. Generally, a larger source leads to shorter onset and peak delays. \\

The results shown in Figs. \ref{fig:onset_times} and \ref{fig:onset_times_vary_source} are now combined and presented in Fig. \ref{fig:onset_times_e_p}. Here we show, in the left panel, the calculated proton onset delay as a function of the calculated electron onset delay. Calculations, shown here as the different symbols, are performed for different source sizes and again compared to measurements from \citet{Richardsonetal2014} (grey data points and green fit to the data). The model results compare well with the measurements and confirm the linear relationship between the electron and proton onset delays. {Due to the east-west asymmetry present in the model, the modelled relationship is not completely linear, but exhibits a closed loop}. In the right panel of Fig. \ref{fig:onset_times_e_p}, the model calculation is repeated, but for a different level of the assumed model background as discussed in the next section. While the best-fit model compares well with the measurements, we argue, again, that the variation of e.g. the background level and accelerator sizes from event-to-event can contribute to explaining the large inter-event variation observed for the onset delays. \\

%%%%%%%%%%%%%%%%%%%%%%%%%%%%%%%%%%%%%%%%%%%%%%%%%%%%%%%%%%%%%%%%%%%%%%
\begin{figure*}
    \centering
\includegraphics[width=0.49\textwidth]{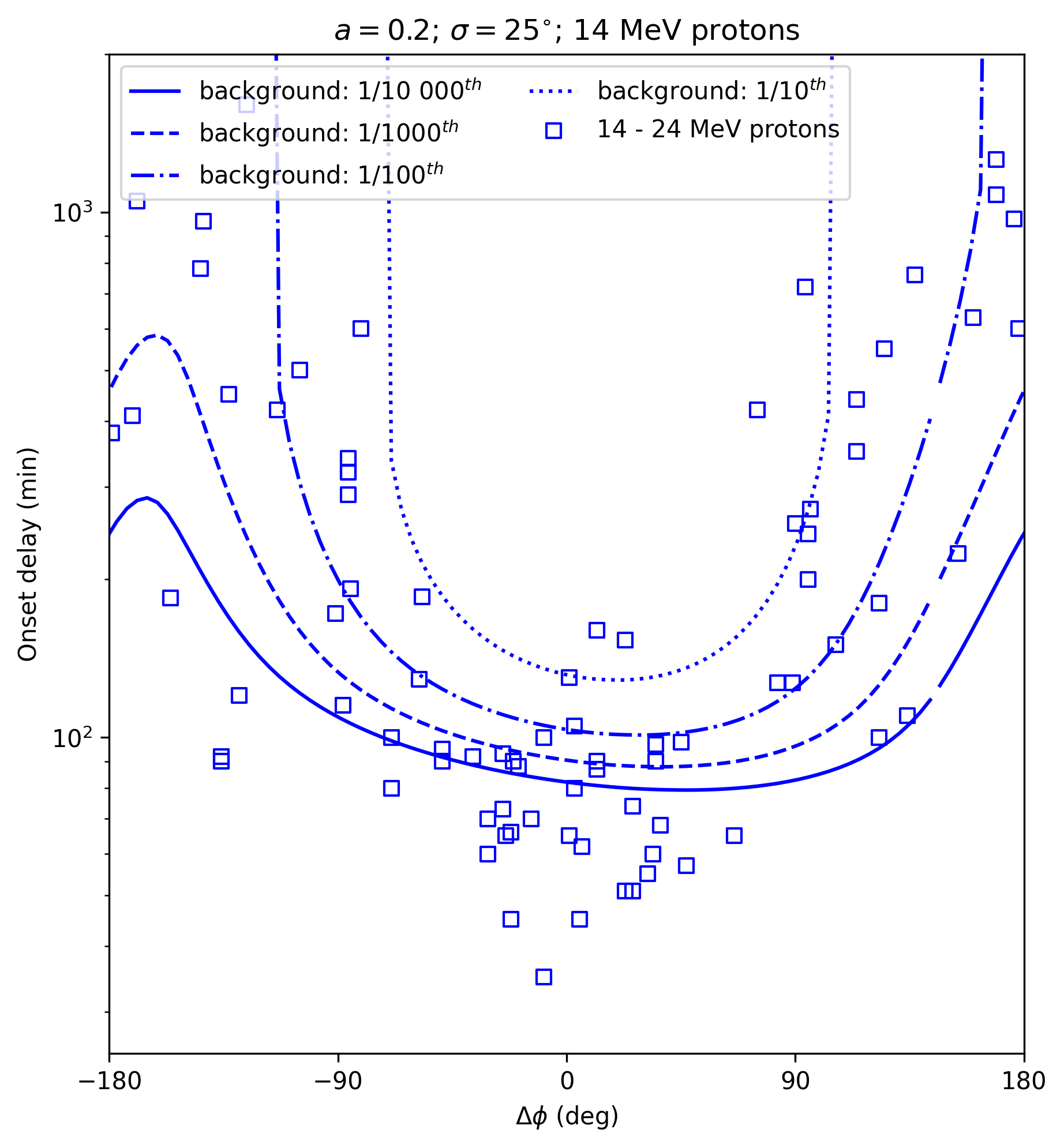}
    \caption{Onset delays for protons, similar to Fig. \ref{fig:onset_times}, but simulations are now performed for different background levels in the model.}
    \label{fig:change_background}
\end{figure*}

\subsection{Caveats in the Simulation of the Onset Delay}
\label{Sec:caveats}

The modelled onset delay is dependent on the assumed background level in the model. The same is true for the observations \citep[see also][]{Laitinenetal2015,Zhaoetal2019}. When the (modelled) differential intensity increases slowly with time, as is the case for observers magnetically disconnected from the source, a lower background will result in a shorter onset delay. This is illustrated in Fig. \ref{fig:change_background} where the background level is changed from 1/10000$^{th}$ of the peak intensity to 1/10$^{th}$ of the peak intensity. The default case is again that of 1/1000$^{th}$ of the peak intensity. It is interesting to note that these different onset delay calculations might also help to explain the observed scatter (inter-event-variations) in the observations: In the model, the same peak intensity is always obtained and the background level is adjusted to simulate the effect of changing background levels whereas for observations of different events, the background of the instrument may remain relatively constant, but the magnitudes (i.e. peak intensity) of the events change. The observed pre-event background from preceding events, if present, may also vary from event to event. In each case, the {\it relative background}, i.e. the peak intensity to background ratio, changes from event to event and can lead to the variation presented in Fig. \ref{fig:change_background}.\\

%%%%%%%%%%%%%%%%%%%%%%%%%%%%%%%%%%%%%%%%%%%%%%%%%%%%%%%%%%%%%%%%%%%%%%
\subsection{Electron and Proton Peak Delays}
\label{Sec:delays}

\begin{figure}[t]
    \centering
\includegraphics[width=0.99\textwidth]{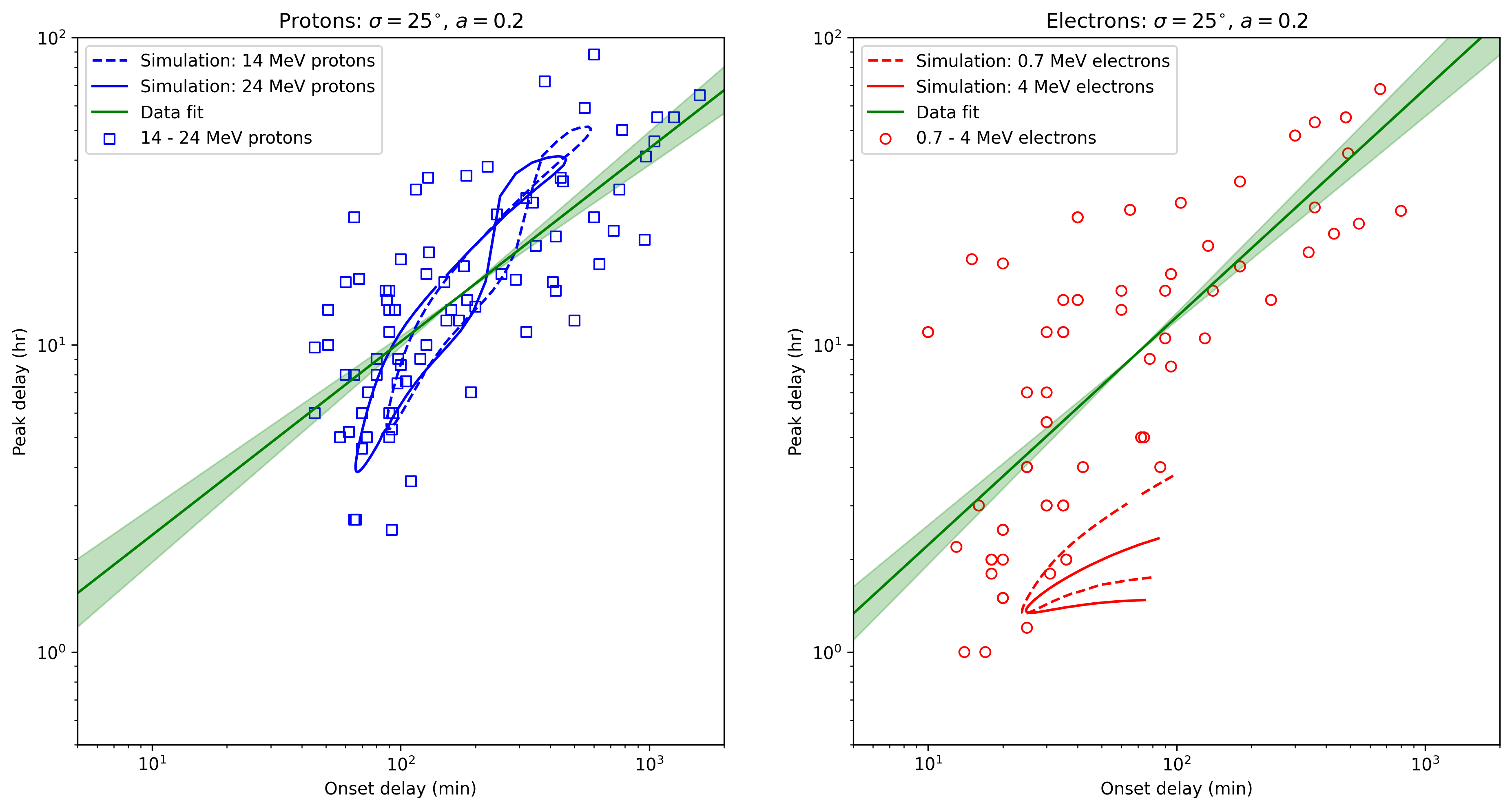}
    \caption{The modelled peak delay as a function of the onset delay for protons (left panel) and electrons (right panel). The measurements are again taken from \citet{Richardsonetal2014}. {Note that, when no model values are shown, the model intensity did not cross the background level at that longitude.}}
    \label{fig:onset_peak_delays}
\end{figure}

As the last part of our work we investigate the modelled relationship between the peak and onset delays. Fig. \ref{fig:onset_peak_delays} shows these simulation results for protons (blue in left panel) and electrons (red in right panel) for different particle energies (different lines). These calculations are again compared to the observations of \citet{Richardsonetal2014}. For protons, the model reproduces the observed linear dependence very well, although these curves are non-linear: Due to the small asymmetry introduced by the Parker HMF, the onset time is slightly shorter at western longitudes and slightly longer at eastern longitudes. On these figures this manifests of this ``figure of eight" shape. These results suggest that both the onset delay and peak delay is at a minimum at the best magnetic connection between the source and observer while both increase for increasingly poor connectivity. For electrons, however, the calculated peak delays are much shorter than the values derived from observations. At present it is not clear why this is the case, although we speculate on possible reasons for this discrepancy in the next section.\\

%%%%%%%%%%%%%%%%%%%%%%%%%%%%%%%%%%%%%%%%%%%%%%%%%%%%%%%%%%%%%%%%%%%%%%
\section{Discussion}

Using transport coefficients calculated from first principles, we are able to simultaneously reproduce the observed longitudinal intensity distribution of 14 - 24 MeV protons and 0.7 - 4 MeV electrons. This is done by assuming an {\it average} source size of $\sigma = 25^{\circ}$. Of course, the measurements show a lot of inter-event variation which can be explained by different interplanetary transport conditions, varying source sizes and, of course, varying levels of peak intensity relative to the background. \\

When perpendicular diffusion is included in the model, we note a systematic shift of the peak intensity to the west of the best magnetic connection due to the geometry of the Parker HMF. Interestingly, a similar shift is seen in both the measurements, but also in previous simulation work by other authors, e.g. \citet{HeWan2015}.\\

Our modelling approach can simultaneously reproduce SEP electron and proton onset delays by assuming the same particle source. When interplanetary transport is included, we do not need to invoke different particle sources expanding at different rates to explain the \citet{Richardsonetal2014} observations. The shortest onset delays are noted at the best magnetic connection, while the delay increases away from best connection due to the slow perpendicular transport process that transports particles across field lines. In fact, we are also able to reproduce the linear relationship observed between electron and proton onset delays by assuming the FLRW process of perpendicular diffusion, where the diffusion coefficient scales linearly with the particle speed. Electrons, being much more mobile than protons at these energies, therefore move faster both along and perpendicular to the mean field. \\

Our results therefore point towards MeV electrons and protons having a common accelerator or source. The transport model itself cannot distinguish between a shock-related or flare-related process. However, it is clear that the model requires both an extended source ($\sigma \sim 25^{\circ}$) and some level of perpendicular diffusion to explain the observed broadness of the SEP events. Such an extended source could point towards a CME-related source. On the other hand, \citet{Dresingetal2016} find no evidence of interplanetary shock acceleration for $\sim 100$ keV electrons, while a comparison with remote sensing observations seem to suggest a flare association for these low energy electrons \citep[e.g.][]{Dresingetal2021}. In previous modelling work we found that 100 keV electron observations can be reproduced, on average, by assuming a compact source with $\sigma \sim 5^{\circ}$. This likely points to a compact flaring source being the main accelerator of electrons at these low energies. Our present and previous simulation results are therefore consistent with the findings of \citet{Dresingetal2022}: Low energy ($\sim$100s of keV) electrons are most likely related to acceleration in compact flaring regions, while $\sim$MeV electrons are most likely produced in more extended CME-driven shock regions. Additionally, our results suggest that $\sim$MeV electrons and $\sim 10$ MeV protons are produced at the same source.\\

One possible deficiency of our present modelling approach is the inability to reproduce the observed peak delays for electrons. The reason for this discrepancy is presently not known. While this could be due to an incorrect level of pitch-angle diffusion included in the model (more diffusion generally leads to longer peak delays), this is unlikely as the amount of pitch-angle scattering needed to increase the peak delay by an order of magnitude would also increase the onset delay with a similar amount. This would be inconsistent with the observations. These longer peak delays are more likely related to an extended injection of SEP electrons into the interplanetary medium due to either particle trapping or particle re-acceleration at the expanding CME shock; {an effect not captured by our current injection profile}. This discrepancy continues to be a topic of further investigation.\\

%%%%%%%%%%%%%%%%%%%%%%%%%%%%%%%%%%%%%%%%%%%%%%%%%%%%%%%%%%%%%%
\acknowledgments

This work is based on the research supported in part by the National Research Foundation of South Africa (NRF grant numbers SRUG220322419 and RA170929263913). Opinions expressed and conclusions arrived at are those of the authors and are not necessarily to be attributed to the NRF. The responsibility of the contents of this work is with the authors. N.D.\ is grateful for support by the Academy of Finland (SHOCKSEE, grant No.\ 346902). IGR acknowledges support from NASA programs NNH19ZDA001N-LWS, NNH19ZDA001N-HSR, and from the STEREO mission. Figures prepared with Matplotlib \citep{Hunter:2007} and certain calculations done with NumPy \citep{Harrisetal2020}.

%=----------------------------------------------------------------------

%%%%%%%%%%%%%%%%%%%%%%%%%%%%%%%%%%%%%%%%%%%%%%%%%%%%%%%%%%%%%%%%%%%%%%
\bibliography{ref}

\end{document}